\documentclass[floatfix,10pt,amssymb,aps,prmaterials,reprint,showpacs,a4paper]{revtex4-2}
\usepackage{amsmath}   
\usepackage{graphicx}  
\usepackage{epstopdf}  
\usepackage{subfigure} 

\newcommand{\U}{$U_{\mathrm{eff}}$}

\renewcommand{\vec}{\mathbf}
\newcommand{\uvec}[1]{\mathbf{\hat{#1}}}

\usepackage[usenames,dvipsnames]{color} 
\usepackage{url}       
\usepackage{hyperref}  
\hypersetup{colorlinks,citecolor=Blue,linkcolor=Blue,urlcolor=Blue}

\begin{document}
\title{Lattice Distortions and Magnetic Interactions in Single-Layer VOCl}
\author{Mohammad Amirabbasi}
\affiliation{Independent Research Center, Shahrood, Iran}
\author{Marcus Ekholm}\email {marcus.ekholm@liu.se}
\affiliation{Link\"oping University, SE-581 83 Link\"oping, Sweden}
\date{\today}

\begin{abstract}
\noindent Atomically thin layers exfoliated from magnetic van der Waals layered materials are currently of high interest in solid state physics.
VOCl is a quasi-two-dimensional layered antiferromagnet which was recently synthesized in monolayer form.
Previous theoretical studies have assumed the high-temperature orthorhombic lattice symmetry also in the low temperature range, where the bulk system is known to be monoclinic due to a strong magnetoelastic coupling.
We demonstrate from \textit{ab-initio} calulations that this monoclinic distortion is prevalent also in monolayers, which is in line with recent experimental indications of monoclinic symmetry.
Our calculations also show that competing ferromagnetic and antiferromagnetic interactions give rise a frustrated two-fold magnetic superstructure where higher-order magnetic interactions play a key role to stabilize the observed magnetic ground state.
\end{abstract}

 \maketitle
\section{Introduction}
\label{sec:introduction}
The recent discovery of spontaneous long-range ferromagnetic order in the two-dimensional (2D) material 
CrI$_{3}$ \cite{huang2017layer} has lead to a surge in the search for such materials by experiments and theoretical calculations alike. 
Stable long-range ordering in low dimension that prospectively could be combined with various tunable properties make them appealing for next generation spintronics devices and functional materials \cite{han2014graphene,wang18,song19,li19,ahn20,sierra21,kurebayashi22}. 
Yet, the fundamental understanding of magnetic interactions in such 2D magnets is still a developing field in solid state theory \cite{burch18,ke21}, as ferro- or antiferromagnetic order in a 2D spin array with isotropic interactions is forbidden at non-zero temperature by the Mermin-Wagner theorem \cite{Mermin1966}.
The observed long-range ordering is commonly attributed to magnetic anisotropy introducing a spin-wave excitation gap~\cite{gong2017discovery}.
In this pursuit, the magnetic van der Waals (vdW) layered materials receive considerable attention, as the weakly bonded layers may be easily exfoliated, and they can be expected to retain the magnetic properties of the bulk material \cite{gong2017discovery,mounet18,chen22}.

\begin{figure}[hbt!]
    \centering
        \subfigure[]{
    \includegraphics[width=0.75\columnwidth]{fig1-a}
    }
        \subfigure[]{
    \includegraphics[width=0.75\columnwidth]{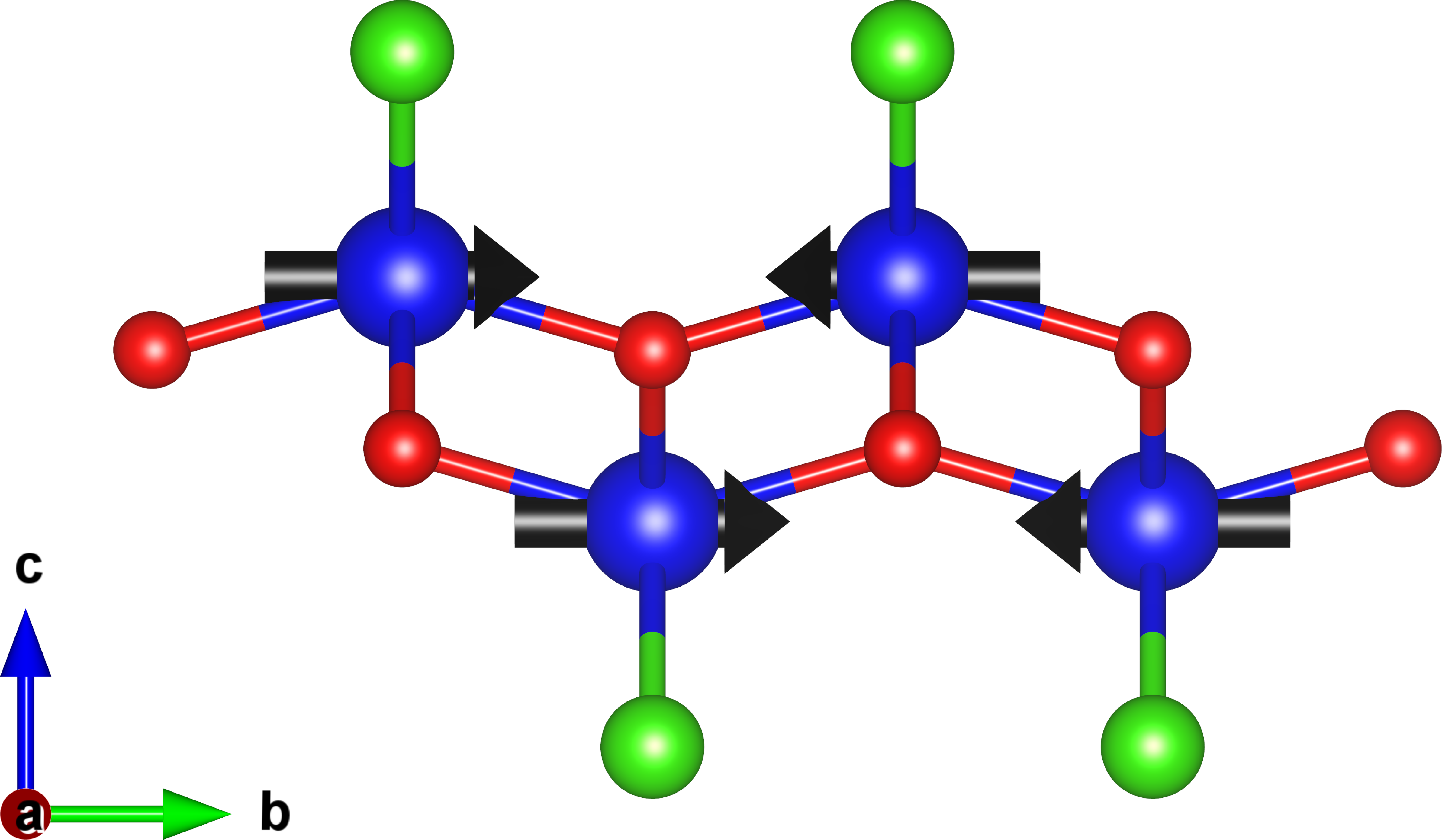}
    }
       \subfigure[]{
    \includegraphics[width=0.82\columnwidth]{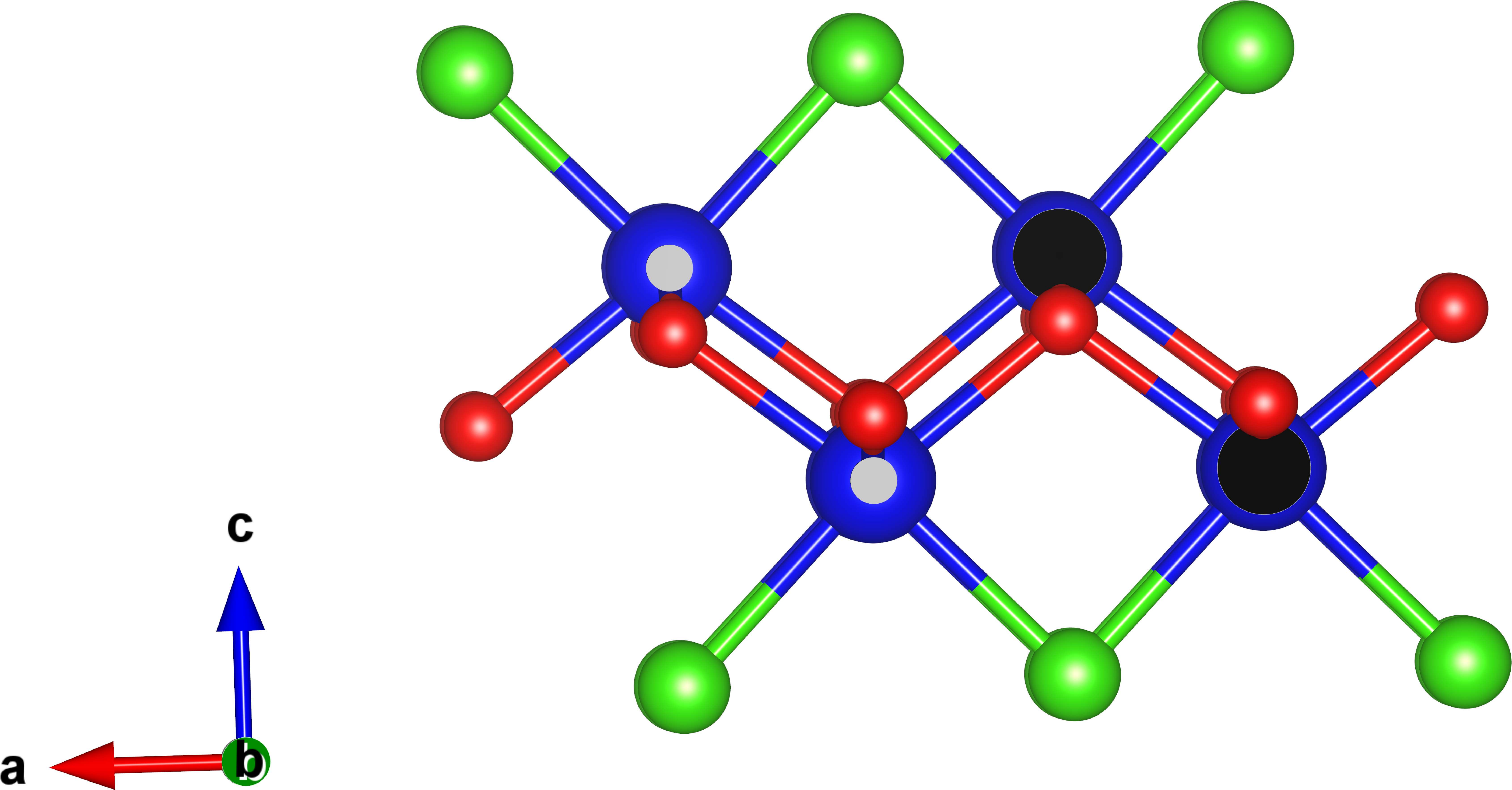}
    }  
      \subfigure[\label{fig:octahedron}]{
                \centering
    \includegraphics[width=0.5\columnwidth]{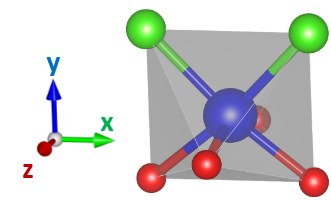}
    }
        \caption{ (a)--(c) Geometry of a single VOCl bilayer viewed along the $\vec{c}$-, $\vec{a}$-, and $\vec{b}$-axes. The green, blue and red spheres denote Cl, V, and O, respectively. The AFM magnetic order corresponding to bulk VOCl is indicated.
(d) The distorted VO$_4$Cl$_2$ octahedron and the local $(x,y,z)$ coordinate system, where $\uvec{x}=-\uvec{a}$, $\uvec{y}=\uvec{c}$, and $\uvec{z}=\uvec{b}$.}
    \label{fig:geometry}
\end{figure}

VOCl is a layered vdW material consisting of V--O bilayers connected by Cl ions on each side, as illustrated in Fig.\ \ref{fig:geometry}.
In its bulk form, these bilayers are separated by a large vdW gap, taking orthorhombic $Pmmn$ symmetry (space group No.\ 58) at ambient conditions \cite{suppmat}.
The crystal structure is common to all the so-called transition metal oxychlorides, $M$OCl, where $M\in \lbrace \mathrm{Ti},\mathrm{V},\mathrm{Cr},\mathrm{Fe}\rbrace$.

At room temperature, bulk VOCl is a paramagnetic insulator, but a twofold antiferromagnetic (AFM) superstructure develops below the N\'eel temperature, $T_N\approx 80$ K \cite{PhysRevB.79.104425,schonleber09}.
The large vdW gap makes VOCl suitable for intercalation applications, and it is currently being considered for novel transistors \cite{zhu21} and battery architectures, with a demonstrated stability to air exposure and cyclic ion shuttling \cite{Zhao2013, Gao_2022, gao2016vocl}.

Single crystals of VOCl with a thickness of only a few atomic layers were first synthesized by Wang \textit{et al}.\ \cite{wang2020atomic}, and were shown to retain the crystal symmetry of the bulk form at room temperature.
However, detailed measurements of the magnetic order of single-layers are challenging and scant.
An \textit{ab-initio} study by Marouche \textit{et al}.\ \cite{mahrouche2021electronic} found ferromagnetic ordering to be the most favorable configuration on a single bilayer with orthorhombic symmetry.
Subsequent theoretical studies \cite{feng21,Li2022} suggested that the corresponding AFM configuration observed in bulk VOCl (see Fig.\ \ref{fig:geometry}) would constitute the magnetic ground state of the single-layers.

Theoretical studies have so far assumed the orthorhombic lattice structure experimentally observed at room temperature \cite{Li2022,mahrouche2021electronic}
The system is then highly frustrated, as each V ion is connected to two V$^{\uparrow}$ and two V$^{\downarrow}$ ions.
Nevertheless, in bulk VOCl, the development of magnetic order below $T_N$ is accompanied by a monoclinic distortion of the crystal structure, lowering the symmetry to $P_{2}/n$ (space group No.\ 13)\cite{PhysRevB.79.104425, schonleber09}.
\textit{Ab-initio} calculations have shown that this distortion is related to magnetoelastic coupling, reducing the V$^{\uparrow}$--V$^{\downarrow}$ distance \cite{Ekholm_2019}.
This monoclinic distortion appears to lift the apparent magnetic frustration that would prevail for the orthorhombic lattice, as the bonds, and the exchange interactions, would be equivalent by symmetry \cite{PhysRevB.79.104425}.
Indeed, recent low temperature measurements on VOCl single-layers have inferred a monoclinic lattice symmetry, although the detailed lattice geometry and magnetic properties require further investigation \cite{villalpando22}.

In this study, we perform structural relaxation of VOCl monolayers by density functional theory (DFT) \cite{dft1,dft2} calculations to show that the AFM configuration leads to a distortion of the lattice that is  completely analogous to the monoclinic distortion of bulk VOCl. 
Assuming this lower lattice symmetry, we derive a magnetic Hamiltonian to study the role of exchange interactions, single-ion anisotropy and the Dzyaloshinskii–Moriya (DM) \cite{moriya1960,DZYALOSHINSKY1958241} interaction in the monoclinic phase.
Monte Carlo simulations recover a $T_N$ comparable to the bulk form.
Our study shows that, counterintuitively, the non-equivalent nearest and next-nearest neighbor interactions are both ferromagnetic; the AFM configuration is due to more long-ranged exchange interactions.
This shows that the system remains frustrated even in the monoclinic phase.

The paper is structured as follows. In Section~\ref{sec:methodology} we provide details of the electronic structure calculations and Monte Carlo simulations.
In Section~\ref{sec:results} we first report on the structural optimization and magnetic order. 
We then describe the electronic structure before detailing the magnetic interactions.
Finally, in Section~\ref{sec:conclusion} we discuss the implications of our results for VOCl single-layers and in the broader context of magnetic vdW layered materials.
\section{Computational details}
\label{sec:methodology} 
Calculations were performed with the Quantum Espresso~\cite{Giannozzi_2009, Giannozzi_2017} code using the GBRV ultra-soft pseudo-potentials~\cite{Vanderbilt2014}, and the all-electron FLEUR \cite{fleur} code.
In Quantum Espresso calculations, we used the cutoffs 50 Ry and 550 Ry when expanding wave functions and charge density in plane waves, respectively. 

In FLEUR-based calculations, the wave function expansion cut-off in the interstitial region was set to $k_{\mathrm{max}}=4.2$ a.u.$^{-1}$. 
The muffin-tin radius of V, Cl, and O atoms were set to 2.28, 2.13, and 1.29 a.u., respectively. 
We have included the $3s$ and $3p$ V-orbitals as semicore states.

For the exchange-correlation energy functional, we have employed the Perdew-Burke-Ernzerhof (PBE) parametrization of the generalized gradient approximation (GGA)~\cite{Perdew1996} with the on-site Coulomb repulsion (DFT+U)~\cite{LDA-U2,Cococcioni2005} applied to the V-$3d$ orbitals.
In FLEUR calculations, the on-site Hund's exchange $J$-parameter was set to $J=1$ eV \cite{vaugier12}, and the on-site Coulomb repulsion, $U$, was varied.
In Quantum Espresso calculations we used the Dudarev parametrization, which requires only the on-site effective Coulomb repulsion, \U$=U-J$ \cite{LDA-U3}.

Magnetic interactions were obtained by fitting a model Hamiltonian to total energy calculations for various magnetic configurations, as described in the Supplemental Material \cite{suppmat}.
To simulate an isolated bilayer, we increased the $c$ lattice parameter to over 30 \AA.
For primitive cell calculations (6 atoms), we used a 20$\times$20$\times$1 optimized Monkhorst-Pack \cite{monkhorst76} $k$-mesh.
The AFM structures require a 2$\times$2$\times$1 cell, and we used a 10$\times$10$\times$1 Monkhorst-Pack $k$-point mesh.

Monte Carlo simulations were performed for a simulation cell containing 12800 spins, using the replica exchange method~\cite{Hukushima1996}.
We performed $2\times10^{6}$ steps for each spin at each temperature. 
To reduce correlation between successive data, statistics were collected every 10 Monte Carlo steps.
Figures of the crystal structures were created with the VESTA software \cite{vesta}.

\section{Results and discussion}
\label{sec:results}
\subsection{Crystal structure and magnetic order}
\label{sec:cryst}
Using the DFT+U method we have optimized the crystal structure while adopting the AFM order previously established for the bulk (see Fig.~\ref{fig:geometry}), for various values of the parameter \U$=U-J$.
As a first step, we constrained the lattice symmetry to orthorhombic, which yields the lattice constants in Table \ref{tab:ortho}.
These values are in agreement with previous calculations \cite{Li2022,feng21}.
Ref.\ \onlinecite{mahrouche2021electronic} reported similar lattice constants for FM ordering.
\begin{table}[!htp]
  \caption{Optimized lattice constants for a single VOCl layer with enforced orthorhombic symmetry, obtained with various \U-values, along with theoretical literature values.\label{tab:ortho}
  }
  \begin{ruledtabular}
      \begin{tabular}{cccc}
               & \U & $a$   & $b$  \\
               & eV  &  \AA  & \AA \\
               \hline
  This work  & 1.0 & 3.31  & 3.80    \\
  ''  & 2.0 & 3.33  & 3.81     \\
 '' & 5.0 & 3.38  & 3.88     \\

  Ref.\ \onlinecite{feng21} & 4 & 3.38 & 3.89 \\
   Ref.\ \onlinecite{Li2022} & 3.25 & 3.36 & 3.86 \\
 Ref.\ \onlinecite{mahrouche2021electronic} & 2 & 3.341 & 3.843 \\
    \end{tabular}
      \end{ruledtabular}
    \end{table}

Lifting the orthorhombic symmetry constraint of the unit cell we find a monoclinic distortion of the crystal lattice for all considered values of the \U-parameter. 
This is in agreement with the recent experimental results by Villalpando \textit{et al}.\ \cite{villalpando22}, who reported a monoclinic lattice symmetry.
The distortion is due to magnetoelastic coupling and is induced by the two-fold AFM superstructure, which breaks the translational symmetry of FM order.
It is completely analogous to what is seen in the bulk; the V$^{\uparrow}$--V$^{\downarrow}$ distance is decreased at the expense of the V$^{\uparrow}$--V$^{\uparrow}$ distance.

Table \ref{tab:mono} accounts for the monoclinic angle $\gamma$ and the lattice parameters obtained with various \U-values.
A larger \U-value will reduce the monoclinic angle, while expanding the lattice.
In the bulk, the value \U$=2$ eV has been shown to simultaneously reproduce structural, electronic and magnetic properties reasonably well \cite{Ekholm_2019,Gao_2022}.
We have calculated the \U-parameter with density functional perturbation theory (DFPT) \cite{Timrov2018}, which resulted in the value \U=$5.67$ eV for both monolayers as well as the bulk.
Consequently, we have taken \U$=5$ eV as an upper limit while considering \U$=2.0$ eV a reasonable value.

For \U$=2$ eV, the distortion lowers total energy by 2.2 meV / atom, and the difference in V$^{\uparrow}$--V$^{\downarrow}$ and V$^{\uparrow}$--V$^{\uparrow}$ distances is 0.019 \AA.
The values of $a$ and $b$ more or less the same, and they are within $\sim 0.02$ \AA\ of what was previously found for bulk VOCl for the same \U-value \cite{Ekholm_2019}.
These values may in turn be compared to the experimental $a=3.30$ \AA\ and $b=3.78$ \AA\ reported in Ref.\ \onlinecite{wang2020atomic} for single crystal results at room temperature.

    \begin{table}[!htp]
      \caption{Optimized lattice constants and monoclinic angle, $\gamma$, for a VOCl single-layer, obtained with various \U-values.
  \label{tab:mono}
  }
  \begin{ruledtabular}
    \begin{tabular}{c c c c}
 \U & $a$ & $b$ &$\gamma$\\
   eV &  \AA  & \AA & $^\circ$ \\
     \hline
      1.0 & 3.31   & 3.80  & 90.68 \\
 2.0 & 3.33   & 3.81  & 90.53 \\
 5.0 & 3.38   & 3.88  & 90.32 \\
  \end{tabular}
  \end{ruledtabular}
\end{table}

The local V spin magnetic moment is $1.5\mu_\text{B}$ and the orbital moment is $-0.079\mu_\text{B}$, which is quite insensitive to the particular choice of \U.
We find the magnetic easy axis to be along $\vec{b}$, which is agreement with the orthorhombic structure, \cite{Li2022,mahrouche2021electronic,wang2020atomic,PhysRevB.79.104425}, and is also analogous to the bulk \cite{Ekholm_2019}.

These results clearly demonstrate that the magnetoelastic properties seen in bulk VOCl carry over to isolated single-layers as well.
All calculations indicate a monoclinic ground state, which is induced by the AFM magnetic order.
An orthorhombic lattice symmetry would indicate magnetic disorder. 

\begin{figure}[!htp]
    \centering
    \includegraphics[width=1.0\columnwidth]{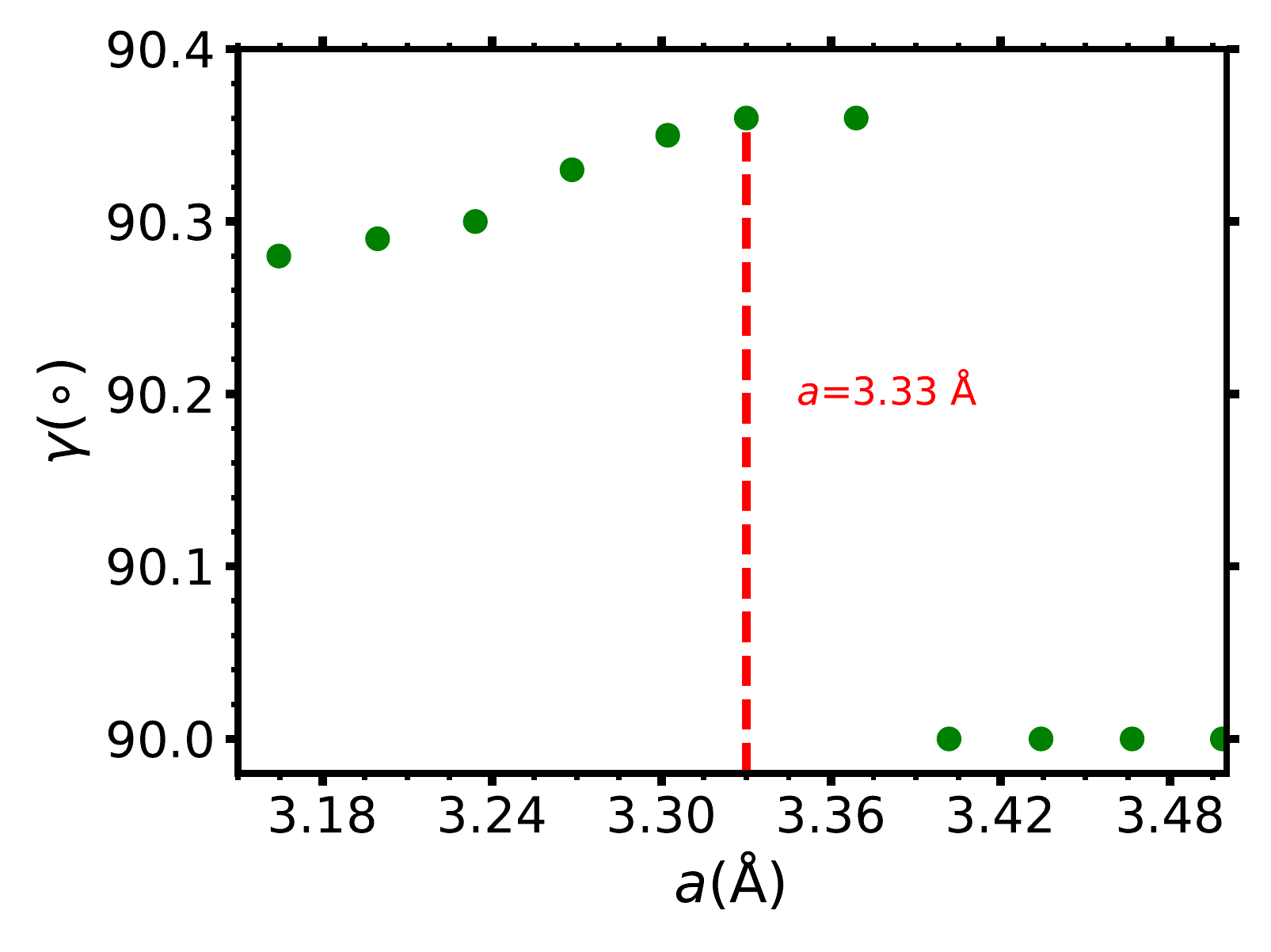}
    \caption{The monoclinic $\gamma$-angle as a function of the lattice constant $a$.
    At each point, the $b/a$ has been optimized.
    }
    \label{fig:str-phase}
\end{figure}

Nevertheless, by expanding the lattice we may recover an AFM orthorhombic structure.
Fig.\ \ref{fig:str-phase} shows the resulting $\gamma$-angle as a function of the $a$ lattice constant.
At each point, the basis coordinates and the $b/a$-ratio has been optimized with \U$=2$ eV.
Above $a=3.37$ \AA\ ($b=3.85$ \AA) the lattice symmetry abruptly changes from monoclinic to orthorhombic, with $\gamma=90.0^\circ$.
As the interionic distances are increased, the magnetic interactions are diminished until there is no elastic energy gain in the distortion, whereupon the lattice changes its symmetry accordingly.
Compressing the lattice has the effect of slightly decreasing the $\gamma$-angle, but no transition is seen in the examined range.

Before discussing further details of the magnetic interactions in Sec.\ \ref{sec:J}, we will outline how the electronic structure of the single-layers compare to the bulk in Sec.\ \ref{sec:elstruct}.

\subsection{Electronic structure}
\label{sec:elstruct}
\begin{figure}[!htp]
    \subfigure[\label{fig:dos_mono_bulk}]{
    \includegraphics[width=1.0\columnwidth]{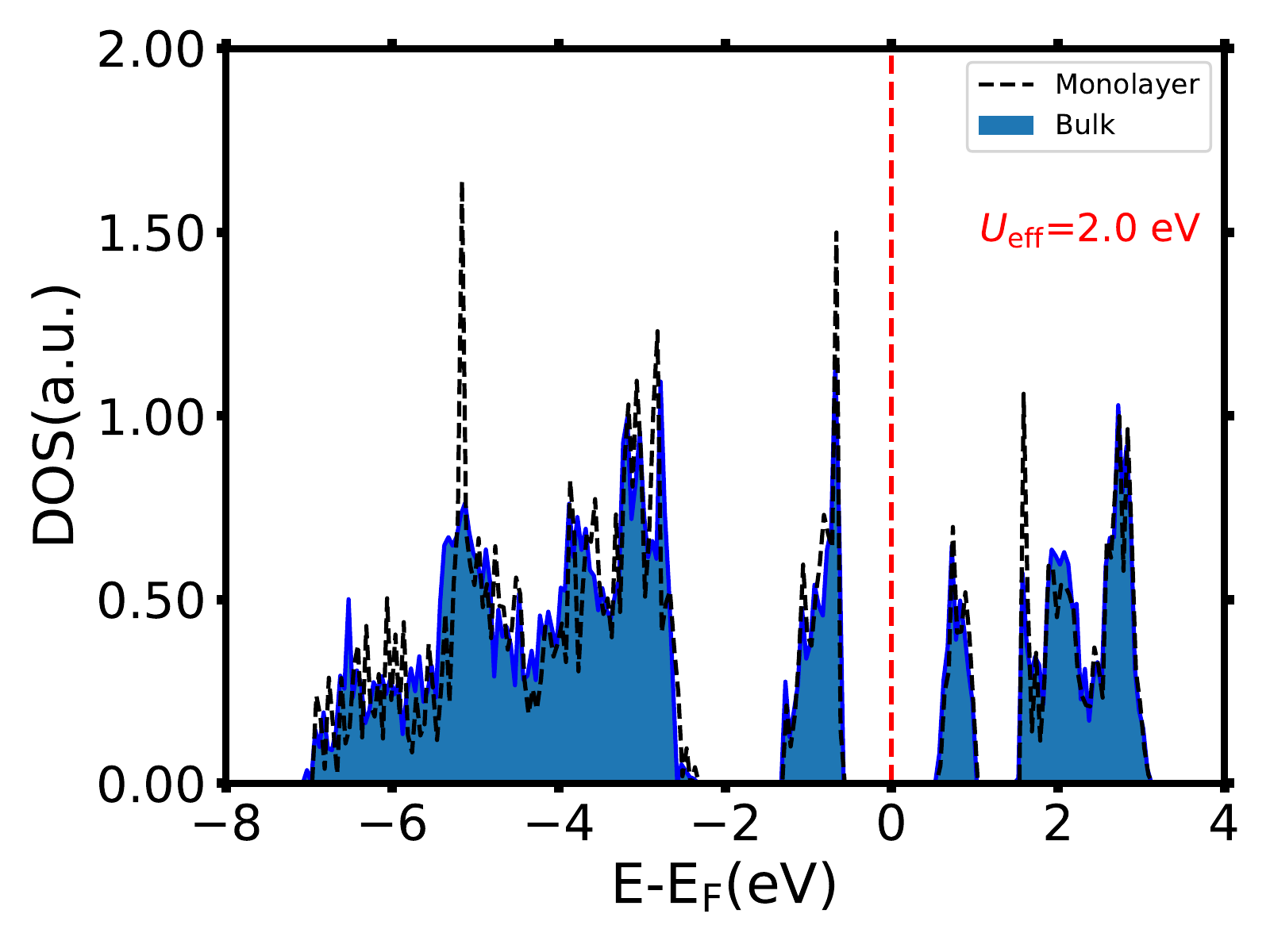}
    }
        \subfigure[\label{fig:band-mono}]{
    \includegraphics[width=1.0\columnwidth]{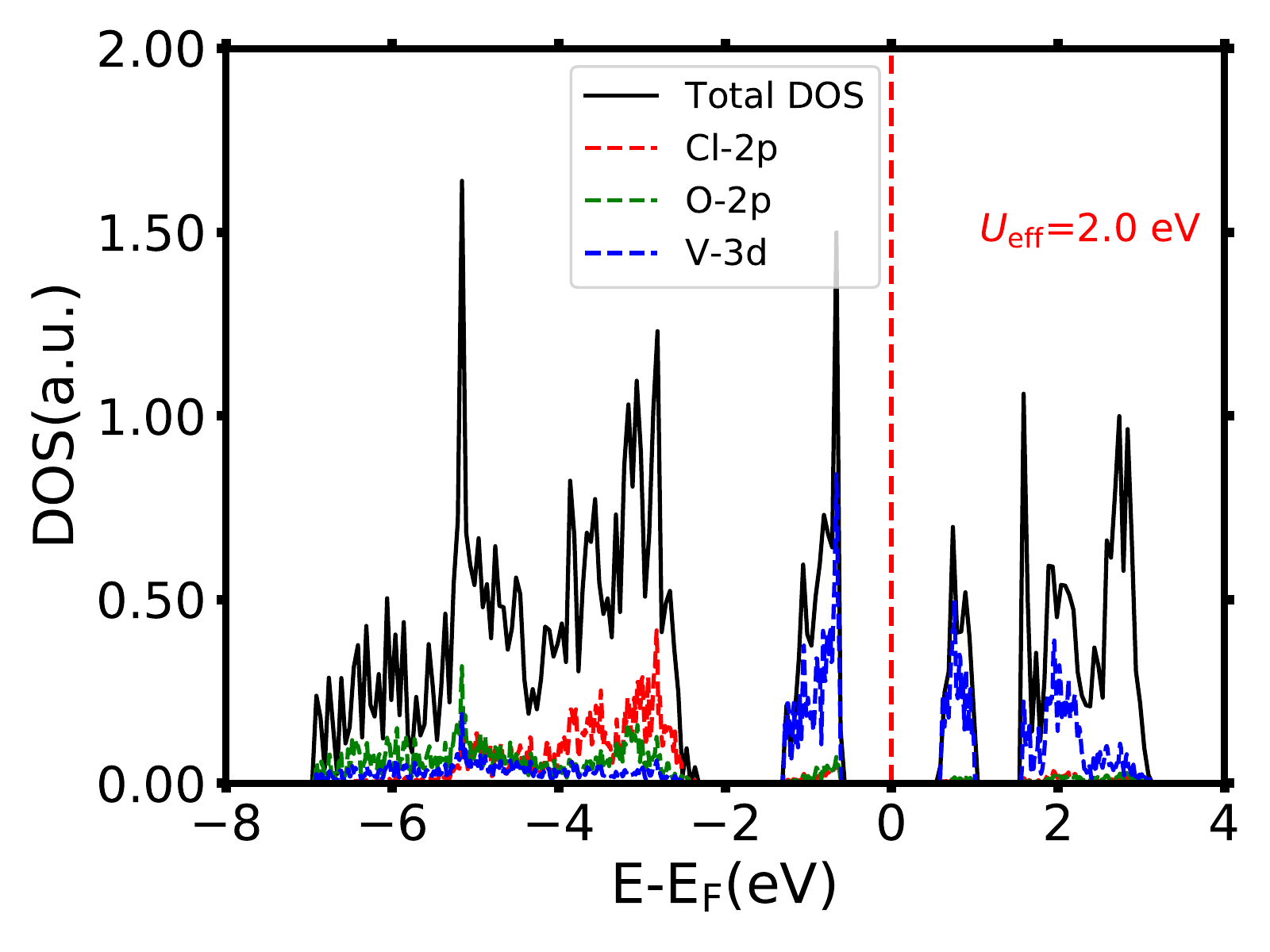}
    }
    \caption{(a) Total DOS of bulk and monolayer VOCl. (b) Site-projected DOS.}
\end{figure}

Fig.\ \ref{fig:dos_mono_bulk} shows the total density of states (DOS) for a VOCl single-layer compared with that of the bulk, calculated with \U$=2$ eV and the AFM order shown in Fig.~\ref{fig:geometry}. 
The electronic structure is very similar in the two cases, with an insulating gap of 1.2 eV.
The similarity underlines the two-dimensional aspects of bulk VOCl, as the single-layers are seen to be quite independent. 
Consequently, VOCl single-layers can be expected to retain the electronic and magnetic properties of the bulk.

In Fig.\ \ref{fig:band-mono} we show the site-projected DOS, indicating that the V-$3d$ electrons dominate the valence states.
These are in turn separated by a gap of 1.05 eV from a manifold of Cl and O states hybridizing with a single V electron.

\begin{figure*}[!htp]
    \centering
    \subfigure[\label{fig:pdos_u2}]{
    \includegraphics[width=1.0\columnwidth]{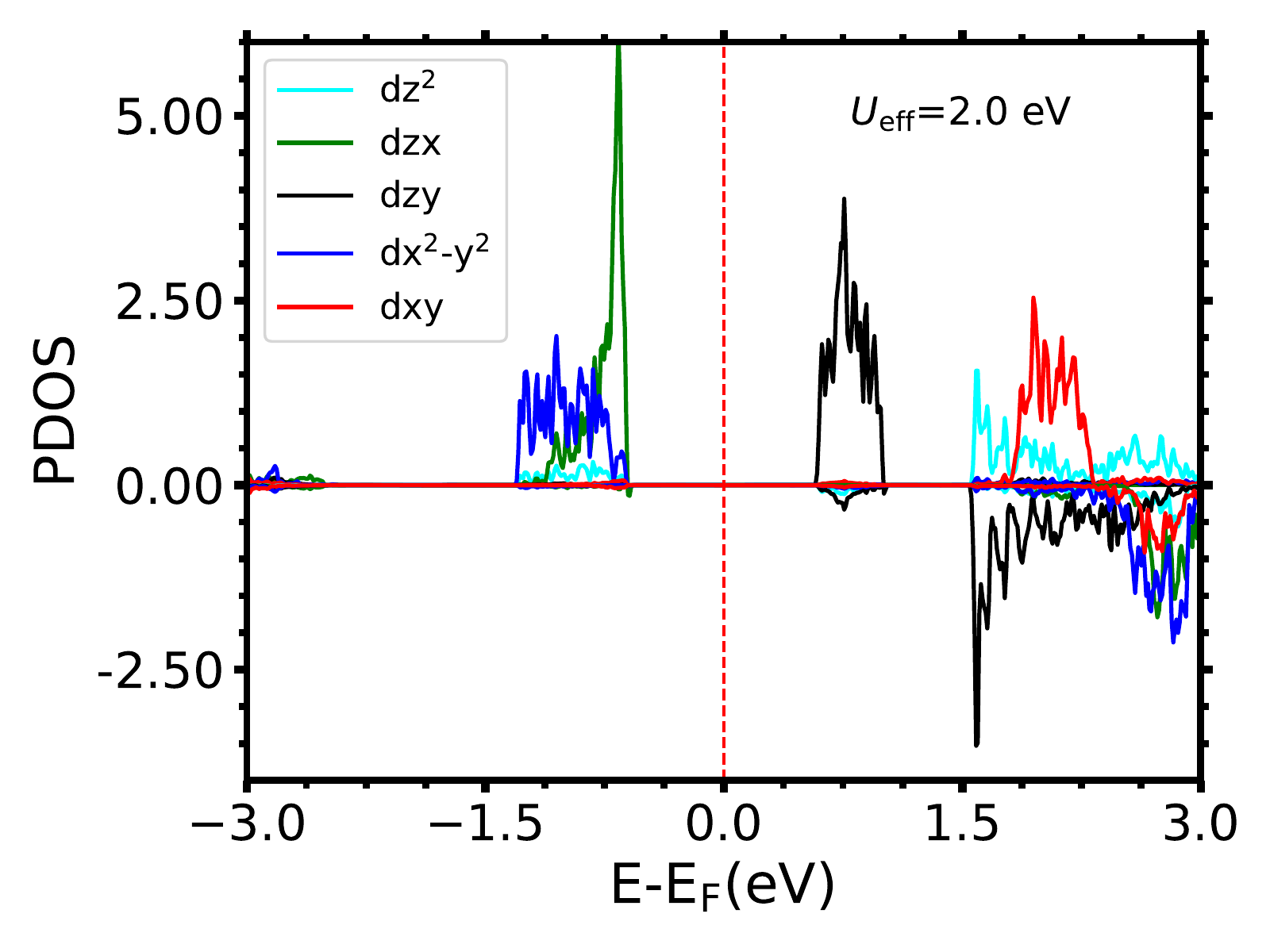}
    }
     \subfigure[\label{fig:bands_u2}]{
        \includegraphics[width=1.0\columnwidth]{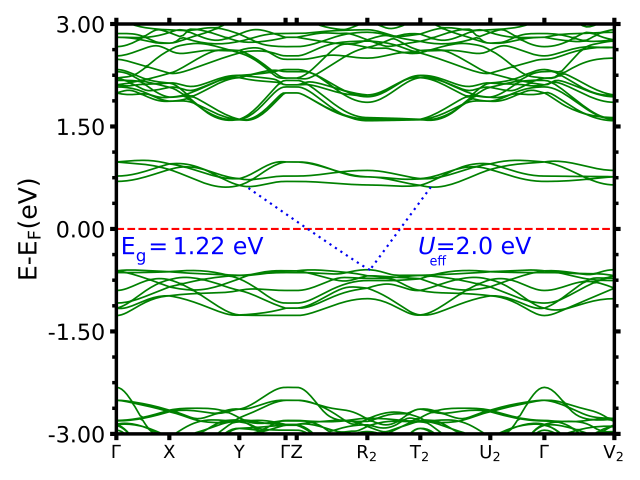}
        }
 \subfigure[\label{fig:pdos_u5}]{
        \includegraphics[width=1.0\columnwidth]{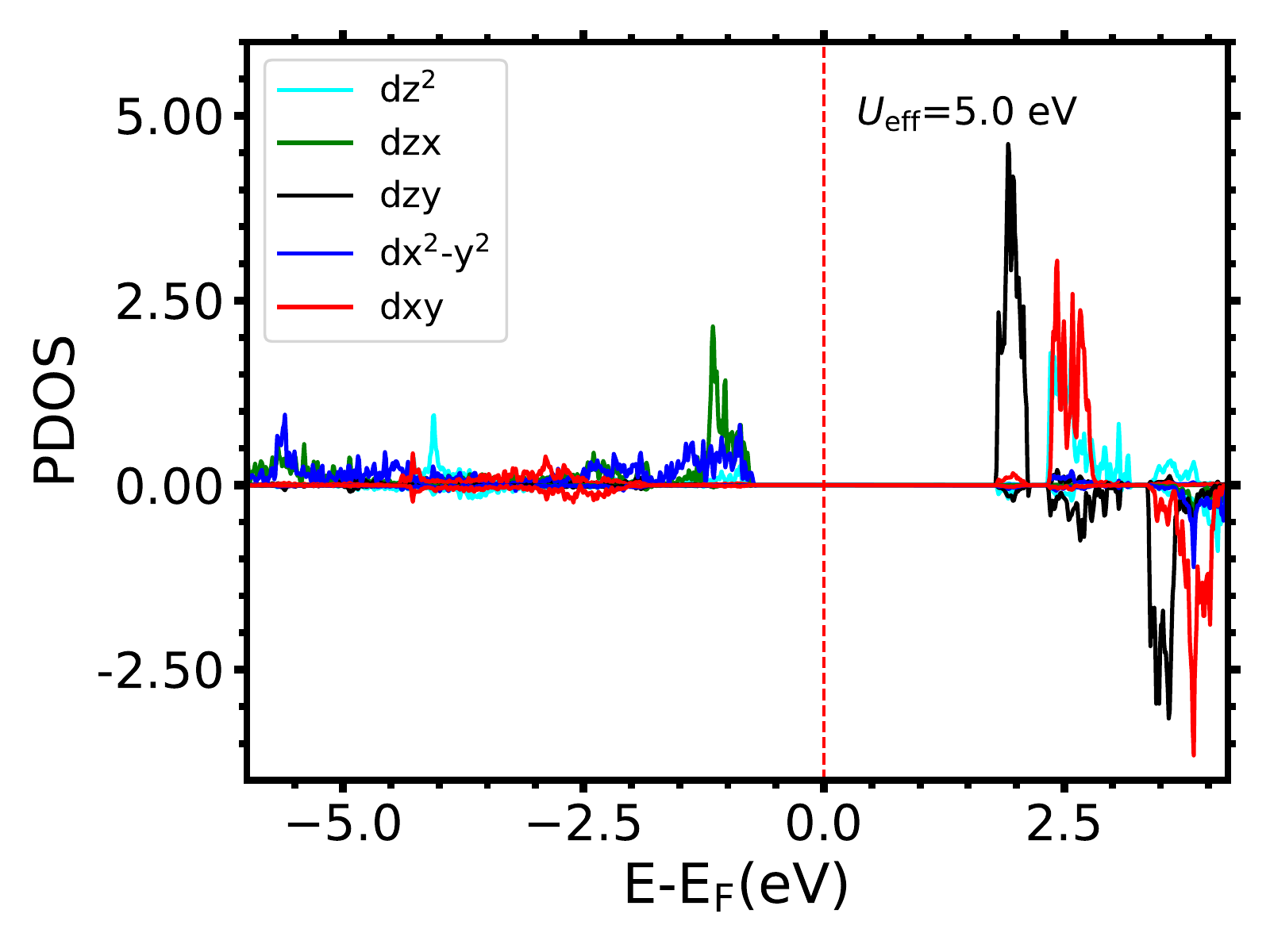}
        }
    \subfigure[\label{fig:bands_u5}]{
        \includegraphics[width=1.0\columnwidth]{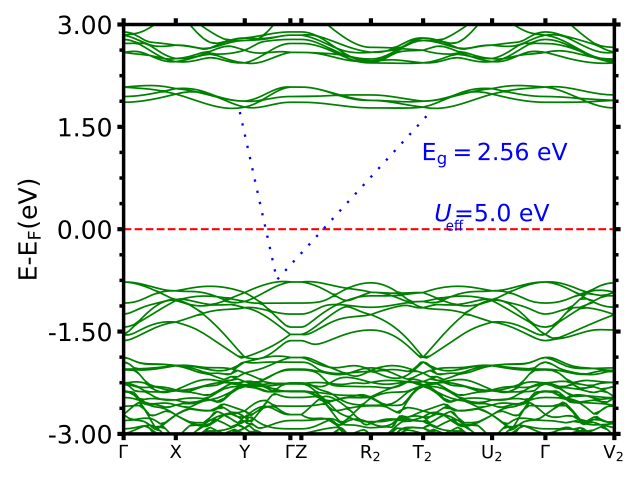}
        }
    \caption{\label{fig:pdos} Orbital-resolved partial DOS (PDOS) and band structure obtained with \U=2 eV and \U=5 eV.
    In (b) and (d), the indirect band gaps are shown by dotted lines. The path are selected based on crystallography~\cite{path1, path2}.
    }
\end{figure*}

The character of the valence V states are seen in Fig.\ \ref{fig:pdos_u2} to be of $d_{zx}$ and $d_{x^2-y^2}$ character.
The lowest unoccupied orbital is of $d_{zy}$ character.
Thus, the degeneracy of the $3d$-levels is completely lifted by the crystal field of the strongly distorted VO$_4$Cl$_2$ octahedron.
Referring to Fig.\ \ref{fig:octahedron}, the $d_{x^2-y^2}$ lobes are directed along the $\vec{a}$ and $\vec{c}$ axes, between the V--O and V--Cl bonds of the $ac$-plane.
The $d_{zx}$ lobes would be most pronounced in the $ab$-plane pointing towards V next-nearest neighbors.

The band structure, shown in Fig.~\ref{fig:bands_u2}, reveals several indirect band gaps of approximately the same size. 
However, this is highly dependent on the \U-value. At the large value of \U$=5$ eV, shown in Fig.\ \ref{fig:pdos_u5} and \ref{fig:bands_u5}, the conduction bands hybridize with the high-binding energy manifold, and the delicate balance between the top and bottom of the conduction and valence bands will shift.
In Ref.\ \cite{feng21} it was reported an indirect $\Gamma$--$X$ gap, which we cannot reproduce.

Having established the crystal symmetry, magnetic order and electronic structure, we will detail the magnetic interactions in the following section.
 
\subsection{Magnetic interactions}
\label{sec:J}
We have derived the coefficients of the following magnetic Hamiltonian \cite{suppmat}:
\begin{eqnarray}
\label{H}
  H &=& - \frac{1}{2}\sum_{i\neq j} J_{ij}(\uvec{S}_i\cdot\uvec{S}_j)+\frac{1}{2} B\sum_{i,j\in \rm nn} (\uvec{S}_i\cdot\uvec{S}_j)^{2} \\ \nonumber
& +&\frac{1}{2} D \sum_{i,j\in \rm nn} \uvec{D}_{ij}\cdot(\uvec{S}_i\times \uvec{S}_j)+\frac{1}{2} \Delta\sum_{i} (\uvec{S}_i\cdot \uvec{d})^{2} \, ,
\end{eqnarray}
where the unit vector $\uvec{S}_i$ denotes the magnetic spin at site $i$.
$J_{ij}$ and $B$ are the bilinear exchange and the nearest neighbor (nn) biquadratic \cite{kartsev20} exchange couplings, respectively.
$\Delta$ denotes the single-ion anisotropy, which is responsible for aligning the spins along the easy axis, $\uvec{d}$.

As the monoclinic distortion removes inversion symmetry, the nearest neighbor DM interaction, $\vec{D}$, may be nonzero.
According to the Moriya rules~\cite{moriya1960}, $\uvec{D}$ should lie in the $ab$-plane. 
Our calculations indeed show that $\uvec{D}$ is directed along the easy axis, $\vec{b}$.
For orthorhombic symmetry, we recover $\vec{D}=\vec{0}$.

The bilinear Heisenberg exchange interactions, $J_{ij}$, deserve particular attention and will be discussed in detail in Sec.\ \ref{sec:exchange} below.
In Sec.\ \ref{sec:spintexture} we will present results from Monte Carlo simulations based on the Hamiltonian  \eqref{H} and discuss the impact of higher order magnetic interactions.

\subsubsection{Bilinear exchange interactions}\label{sec:exchange}
\begin{figure*}[!htp]
    \centering
    \subfigure[\label{fig:Jij}]{
    \includegraphics[width=1.0\columnwidth]{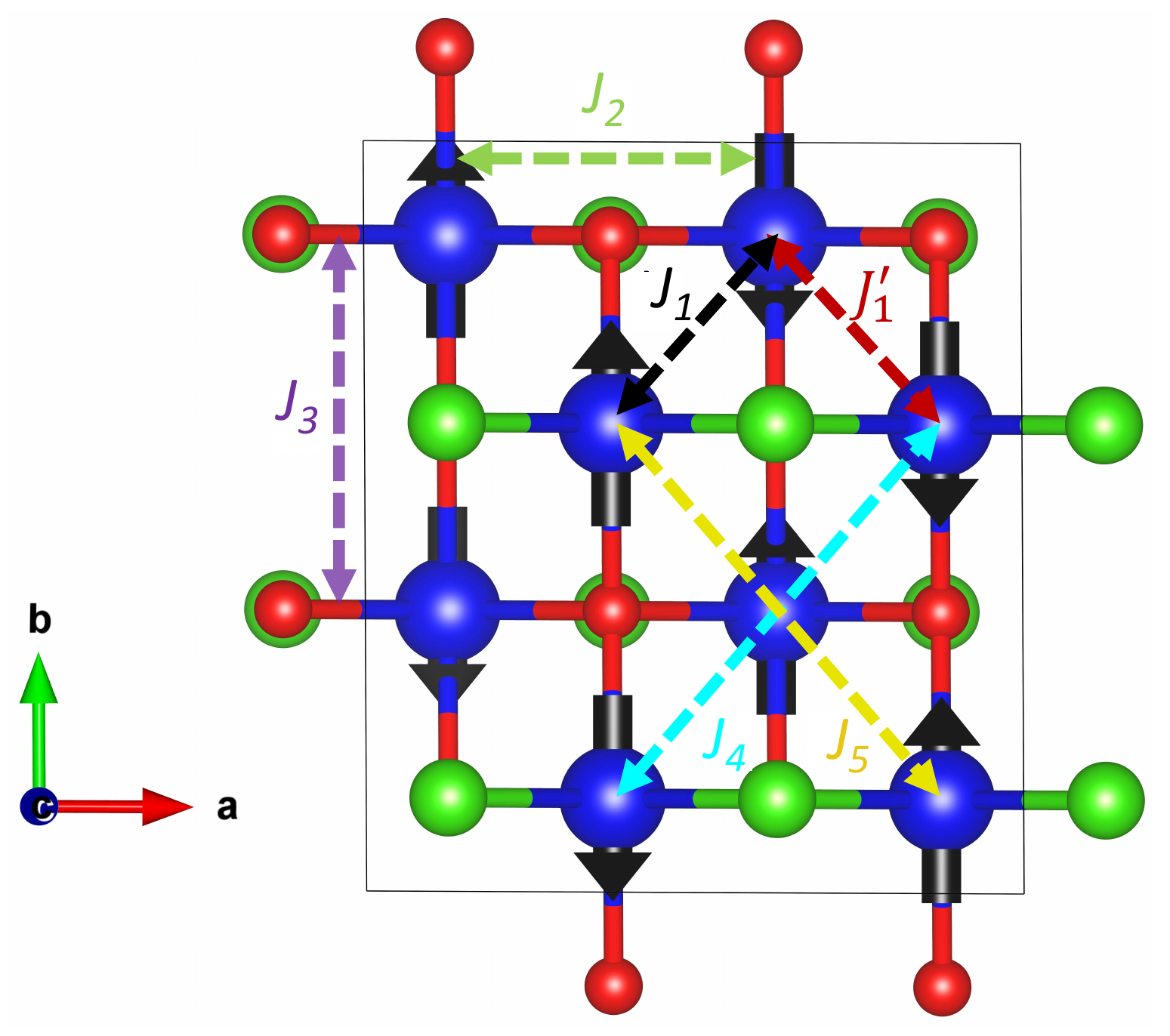}
    }
    \subfigure[\label{fig:Jij2}]{
    \includegraphics[width=0.78\columnwidth]{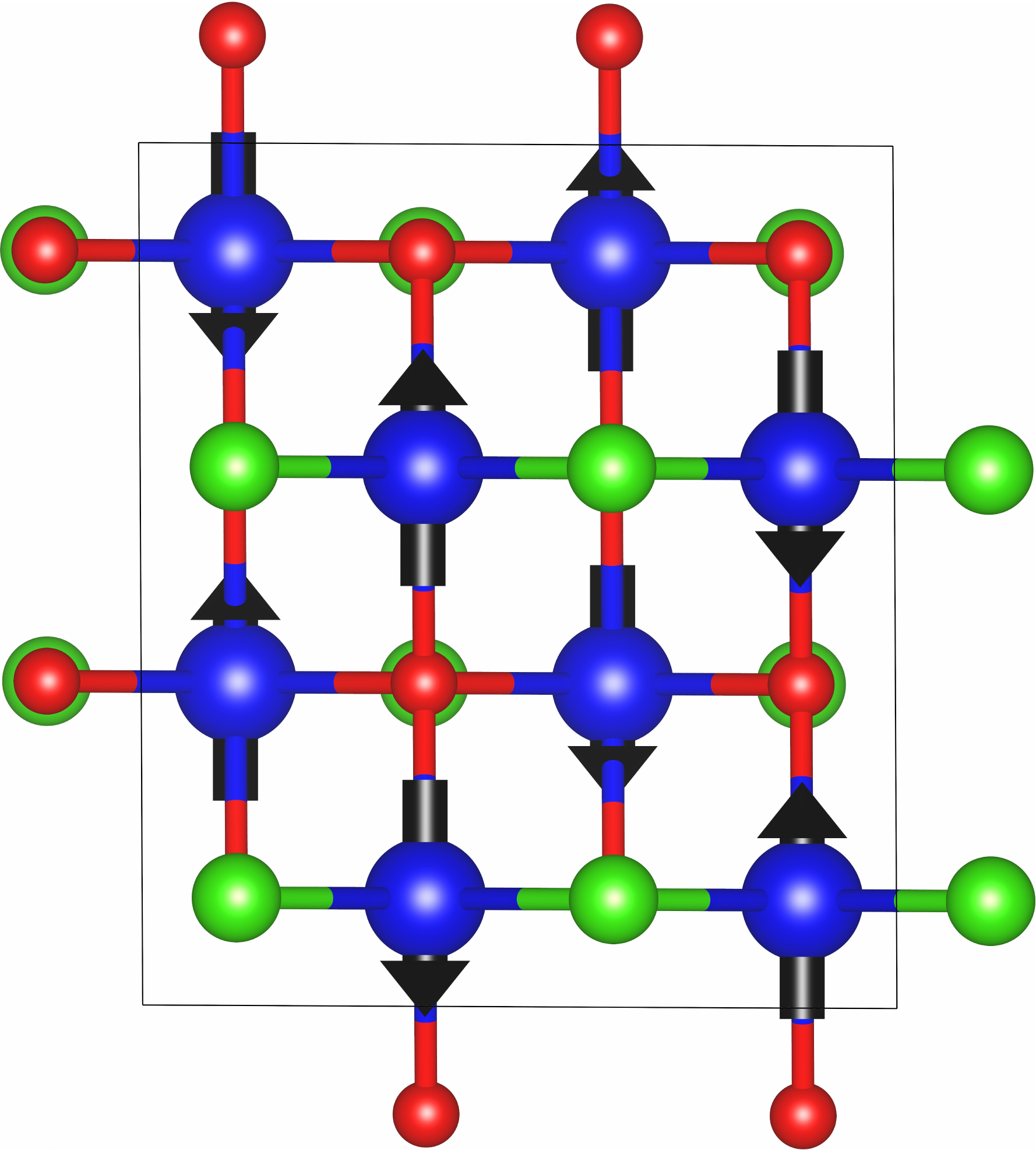}
    }
    \caption{(a) Exchange couplings in a VOCl bilayer. The $J_1$ and $J_1'$ interactions act between V$^\uparrow$--V$^\downarrow$ and V$^\uparrow$--V$^\uparrow$ ions, respectively. The $J_2$ and $J_3$ interactions are between nearest neighbors along $\vec{a}$ and $\vec{b}$.
The magnetic configuration in (a) and (b) are degenerate for orthorhombic symmetry, where $J_1 = J_1'$.
A monoclinic distortion of the lattice reduces the $J_1'$ distance.
    }
    \label{fig:exhange_couplings}
\end{figure*}

In Fig.\ \ref{fig:Jij} we highlight the most relevant Heisenberg exchange-couplings on the VOCl lattice.
We denote the nearest neighbor interaction by $J_1$ and the second nearest neighbor $J_1'$.
In the orthorhombic case, $J_1$ and $J_1'$ are equivalent by symmetry as discussed in Sec.\ \ref{sec:cryst}. 
This means that the spin configuration in Fig.\ \ref{fig:Jij2} is degenerate with that of Fig.\ \ref{fig:Jij}.
For monoclinic symmetry, this degeneracy has been lifted, and the configuration in Fig.\ \ref{fig:Jij} is the ground state.

The $J_2$ and $J_3$ interactions act between ions separated by $\vec{a}$ and $\vec{b}$ (note that $a>b$), while the $J_4$ and $J_5$ act along $\pm \vec{a}\pm \vec{b}$.

\begin{table*}
\begin{ruledtabular}
  \caption{Calculated Heisenberg couplings ($J_1$--$J_6$), biquadratic exchange ($B$), single-ion anisotropy $\Delta$, and DM interaction ($D$) for different \U-parameters in the monoclinic structure. 
Positive (negative) values denote (anti-)ferromagnetic coupling.
  $T_\text{N}$ is the N\'eel temperature obtained from Monte Carlo simulations~\cite{li2021spin, wang2022impacts, illas2000magnetic}.
  \label{tab3}
}
      \begin{tabular}{ccccccccccc}
    $U_\text{eff}$& $J_\text{1}$& $J_\text{1}'$& $J_\text{2}$& $J_\text{3}$& $J_\text{4}$&$J_\text{5}$&$B$&$\Delta$&$D$&$T_\text{N}$\\
    \hline
    eV & \multicolumn{9}{c}{meV}& K \\
    \hline
            1.0 & 4.34  & 6.43  & -9.84 & -7.94 &-0.09 &-0.09&-3.58&-0.13&-0.60&98.67 \\
            2.0 & 3.62  & 4.96  & -6.10 & -6.42 &-0.07 &-0.07&-3.55&-0.15&-0.66&73.33\\
            5.0 & 2.79  & 3.32  & -1.31 & -3.68 &-0.00 &-0.00&-2.88&-0.11&-0.65&32.16 \\
     \end{tabular}
     \end{ruledtabular}
               \end{table*}
               
The obtained magnetic interactions are summarized in Table \ref{tab3}, which shows that $J_4$ and $J_5$ are negligible, and will therefore not be discussed further.
The nearest neighbor interactions, $J_1$ and $J_1'$ are both seen to favor FM order, while the $J_2$ and $J_3$ interactions are AFM, regardless of the \U-parameter.
This means that it is not primarily the nearest neighbor interactions which are responsible for the magnetic ordering, as anticipated in the literature \cite{PhysRevB.79.104425}, but the more distant $J_2$ and $J_3$ interactions.
It also means that the the monoclinic phase is frustrated: each V$^{\uparrow / \downarrow}$ ion has two V$^\uparrow$ and two V$^\downarrow$ nearest neighbors, although all four nearest neighbor exchange interactions actually favor FM alignment.

The nearest-neigbor exchange interactions in bulk VOCl have previously been discussed in terms of a combination of direct exchange mediated by the V $d_{zx}$ orbitals and superexchange involving $d_{x^2-y^2}$ electrons \cite{glawion09}. Indeed, the former indeed point approximately along a line connecting the V--V nearest neighbors, while the latter involves two V--O--V paths. 
The sign of the $J_1$ and $J_1'$ interactions are consistent with the Goodenough–Kanamori–Anderson (GKA) rule~\cite{GKA} for superexchange, applied to the two V--O--V paths: the bond angles are close to 90$^\circ$  (99.5$^\circ$ and 100.4$^\circ$) with the same total bond length, which would favor FM ordering. 

The V--O--V bond angle along $\vec{b}$ is $147.5^\circ$, closer to $180^\circ$ which would favor AFM order for the $J_2$ interaction, as observed.
As noted in Sec.\ \ref{sec:elstruct} the lobes of the $d_{x^2-y^2}$ orbitals point along $\vec{a}$ and $\vec{b}$, which indeed would support the V--O--V hopping path.
However, the $J_3$ interaction along $\vec{a}$ is mediated by a V--Cl--V bond, which forms a $97.3^\circ$ angle, together with the V--O--V $103^\circ$ bond, yet the $J_3$ interaction is AFM, contradicting the GKA rule.

It remains an open question how the superexchange mechanism would work in polyvalent materials, such as VOCl, although attempts have been made to construct a theory for CrOCl and FeOCl \cite{zhang19}.
Although direct overlap may seem unlikely, it cannot be ruled out that the exchange interactions are mediated by a combination of direct and indirect exchange.
Most likely, there is a competition between Pauli exchange, Hund's coupling, and dynamical electron correlation \cite{qing20,jang21}.

Varying the \U-parameter, we also find that interactions are reduced. 
The nearest neighbor interaction $J_1$ is always smaller than $J_1'$, although it has a shorter V--V distance and also corresponds to the smaller 99.5$^\circ$ V--O--V bond. It seems as if the forced AFM order between the V--V nearest neighbors leads to a reduction of the FM exchange interactions.

However, the ratio between the AFM $J_{2,3}$ and the FM $J_1$/$J_1'$ will vary with \U. In particular, the $J_2$ interaction which connects $V$-spins along $\vec{a}$ is affected most strongly.
For \U$=1$ eV, the $J_{2,3}$ interactions dominate $J_1$ and $J_1'$, and $J_2$ is by far the strongest.
For \U$=5$ eV, $J_2$ is instead the weakest interaction, and $J_3$ is comparable to the $J_1$/$J_1'$ interactions.

In the orbital-resolved DOS of Fig.\ \ref{fig:pdos}, the $t_{2g}$ and $e_g$ orbitals are seen to be well separated from the high energy manifold for \U$=2$ eV.
But for \U$=5$ eV, the hybridization of these orbitals is significant.
The band gap has also been effectively doubled, which reduces the hopping tendency to the unoccupied $d_{zy}$ states along V--V bonds in the $bc$-plane.
The dominating effect seems to be the latter, which reduces the hopping of the large-angle V--O--V bond along $b$ responsible for $J_2$. 

It is interesting to compare the exchange interactions on the monoclinic lattice with those of the orthorhombic lattice.
Our calculations \cite{suppmat} (for \U$=2$ eV) yield a smaller FM $J_1$ parameter of 1.27 meV, with comparable AFM $J_{2,3}$.
These results are in qualitative agreement with Ref.\ \cite{Li2022}, although those interactions were derived from a smaller set of magnetic configurations, and the implications of the results were never discussed.
Allowing the lattice to we thus observe an increase in the FM nearest neighbor interactions.

Glawion \textit{et al}.\ \cite{glawion09} considered exchange interactions for the orthorhombic bulk system and also reported AFM interactions along $\mathbf{a}$ and $\mathbf{b}$, but found the sign of $J_1$ to depend on the assumed \U-value.
We do not see this effect in the single-layers, which may be due to an assumed FM state in Ref.\ \onlinecite{glawion09}.
In any case, all theoretical work agrees on competing FM and AFM interactions in the VOCl system, giving rise to frustration.

\subsubsection{Spin texture}\label{sec:spintexture}
\begin{figure}[!htp]
    \centering
    \subfigure[\label{fig:order_parameter}]{
    \includegraphics[width=1.0\columnwidth]{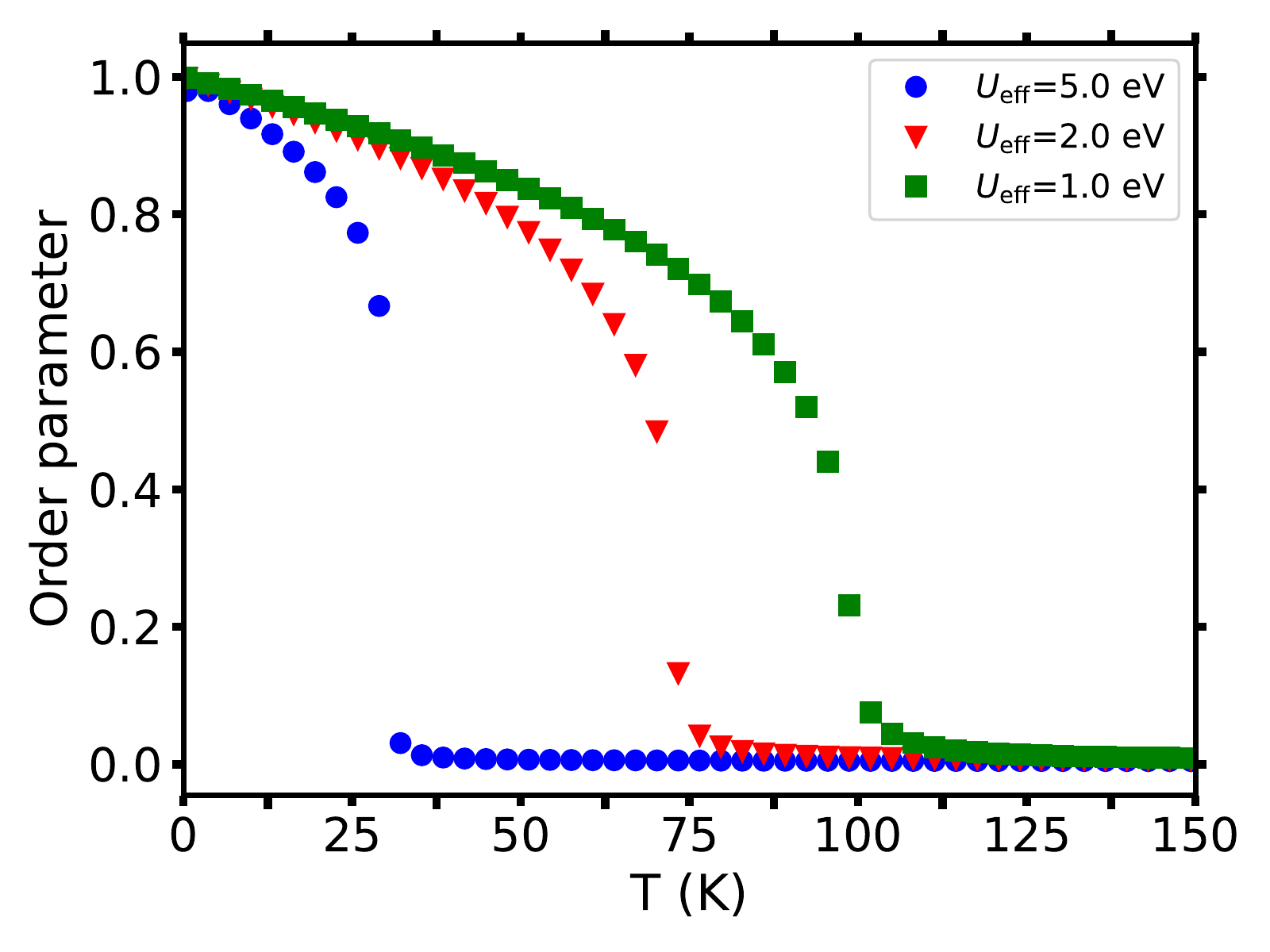}
    }
        \subfigure[\label{fig:specific_heat}]{
    \includegraphics[width=1.0\columnwidth]{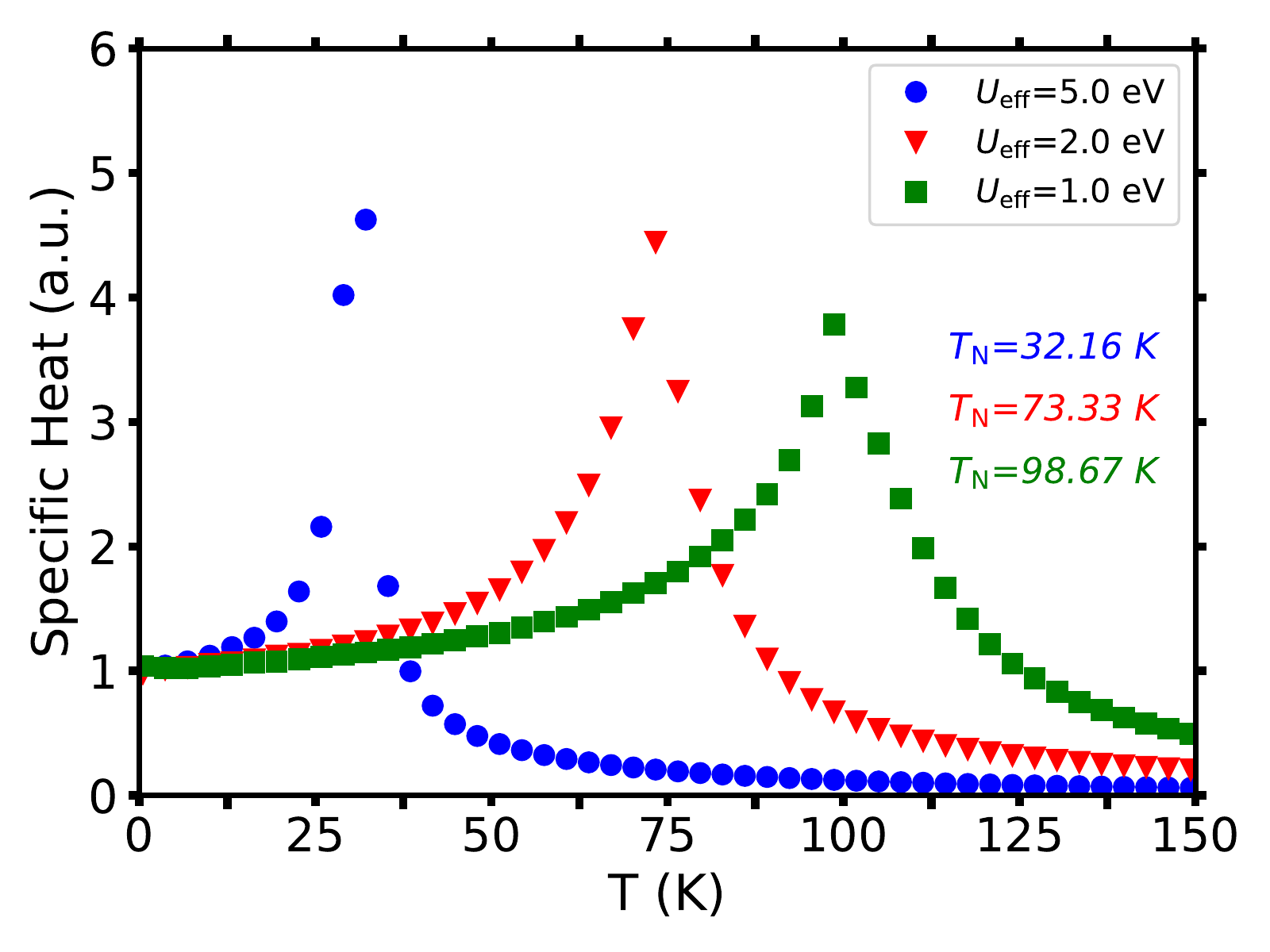}
    }
    \caption{(a) AFM order parameter, $m$, and (b) magnetic (specific) heat capacity as a function of temperature, $T$, for different $U_\text{eff}$ parameters.}
\end{figure}

As a test of the calculated magnetic interactions, we have performed Heisenberg Monte Carlo simulations on the monoclinic lattice.
We define an AFM order parameter as $m=\frac{1}{N}\sum^{N}_{i=1} \uvec{S}_{i}\cdot \uvec{d}_{i}$, where $\uvec{d}_i=\pm \uvec{b}$ is the ideal direction of the spin at site $i$, and $N$ is the total number of spins.
$m=1$ thus corresponds to the AFM ground state and $m=0$ indicates complete disorder.
Fig.\ \ref{fig:order_parameter} shows the order parameter as a function of temperature for various values of \U, which reaches $m=0.98$ at $T=0.5$ K.
The magnetic heat capacity is plotted in Fig.\ \ref{fig:specific_heat} and is seen to reach a finite value in the $T \rightarrow 0$ limit.
The N\'eel temperature, $T_N$, is taken from the divergence of the heat capacity, and is listed in Table~\ref{tab3}. 
For \U$=2$ eV, we obtain $T_N= 70$ K, which is comparable to the experimental value of 80 K for bulk VOCl \cite{WIEDENMANN1984275,PhysRevB.79.104425}.

$T_N$ is mostly determined by the size of the isotropic $J_{ij}$-parameters.
However, the biquadratic exchange interaction, $B$, is comparable with the $J_{ij}$-values and gives a non-negligible contribution to $T_N$.
Neglecting the biquadratic exchange would lower $T_N$ by 13 K for \U$=2$ eV.

The DM-interaction parameter, $D$, is seen in Table~\ref{tab3} to be quite small and only weakly dependent on \U.
It does not give any appreciable contribution to $T_N$.
The negative values of $D$ indicate that the DM interaction tends to make the spins collinear to each other \cite{canals2008ising, elhajal2005ordering}.

\begin{figure}
 \subfigure[\label{fig:spintexture}]{
    \includegraphics[width=0.8\columnwidth]{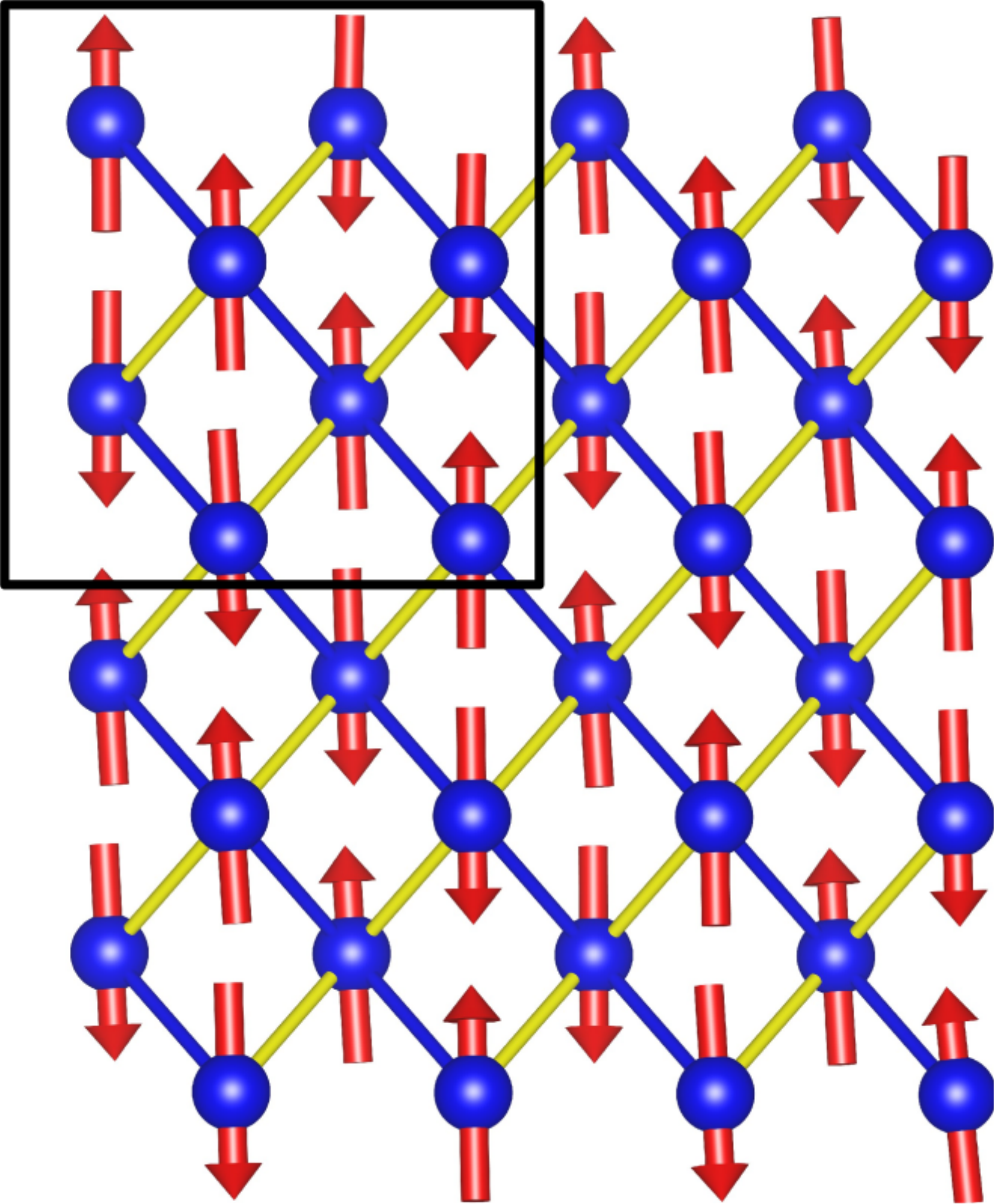}
    }    
 \subfigure[\label{fig:histogram}]{
    \includegraphics[width=1.0\columnwidth]{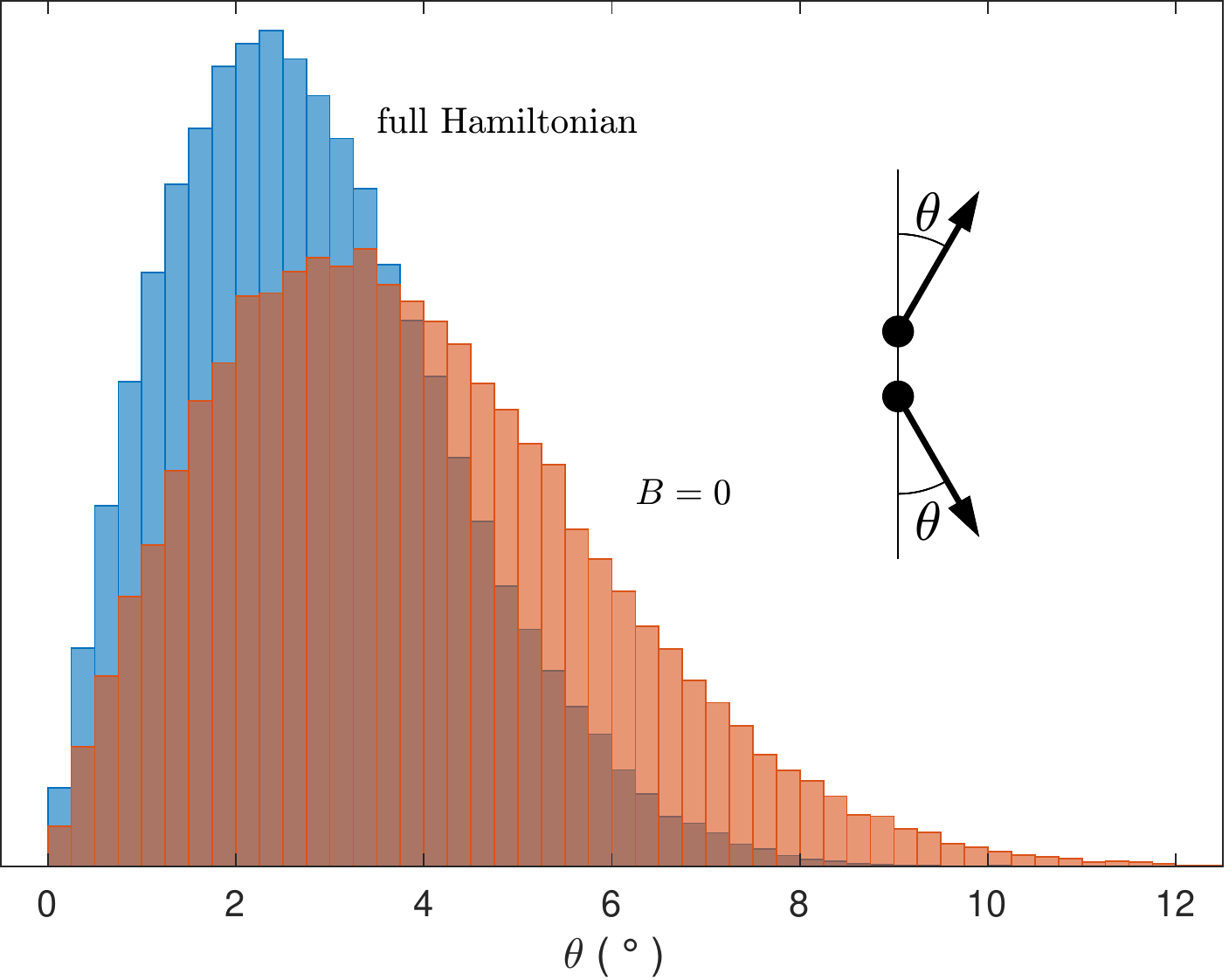}
    }
        \caption{(a) Spin snapshot from Monte Carlo simulations at $T=0.5$ K. The black rectangle indicates a $2\times 2$ magnetic unit cell. Yellow lines highlight nearest-neighbors and blue lines next-nearest neighbors.
    (b) Histogram of the angle between the spins and the global quantization axis, $\theta$, obtained with the full Hamiltonian of Eq.\ \eqref{H} (blue), and setting the biquadratic term $B=0$ (red).}
\end{figure}

Fig.\ \ref{fig:spintexture} shows a snapshot of the spin texture at $0.5$ K on the monoclinic lattice.
Although the correct AFM ground state is reached, we observe slight deviations from collinearity.
Averaging the deviation angle, $\theta$, of the local spin moments from the global quantization axis over all atoms in 100 individual simulation cells, we find $\left\langle \theta \right\rangle = 2.8^\circ$.
The corresponding distribution is shown in Fig.\ \ref{fig:histogram}. 
Apart from $J_{ij}$, the most important term for the alignment of the spins seems to be the biquadratic interaction, $B$.
Removing the biquadratic term by setting $B=0$ leads to $\left\langle \theta \right\rangle = 3.9^\circ$ with a larger variation of $\theta$.

Ref.\ \cite{li2021spin} reported the much smaller value of $T_N=23$ K from Monte Carlo simulations on the orthorhombic lattice. 
In the orthorhombic case we find that the system jumps between the two degenerate magnetic solutions below $T_N$ (see Fig.\ \ref{fig:exhange_couplings}) leading to a non-monotonic dependence of the order parameter with temperature. 
This is interesting, as the monoclinic angle is temperature dependent, pointing at the importance of the spin-lattice coupling, that ideally should be taken into account. 
However, this is beyond the scope of this study, which targets the magnetic ground state.

\section{Summary and Conclusions}
\label{sec:conclusion}
In summary, using DFT+U calculations we have determined structural and magnetic properties of the single VOCl bilayer.
Our PBE+U calculations show that the system undergoes the same monoclinic distortion as previously observed in bulk VOCl \cite{Ekholm_2019,schonleber09,PhysRevB.79.104425}.
The monoclinic AFM phase is magnetically frustrated, as the nearest neighbor interactions are all FM, and the observed AFM order is in fact enforced by longer ranged AFM interactions.
Thus, the monoclinic distortion does not remove the magnetic frustration.
These conclusions are independent of the particular value of the \U-parameter and are in line with recent experimental reports of a monolinic lattice symmetry \cite{villalpando22}.

Together with our calculations of the electronic structure, we conclude that the physical properties of the individual layers of bulk VOCl carry over to single-layers. Nevertheless, it should be remembered that the layers of bulk VOCl are not completely magnetically independent, as they do form a well ordered two-fold magnetic superstructure along $\vec{c}$ as well. 

By means of Monte Carlo simulations we have calculated the N\'eel temperature, which will depend on the \U-value.
With \U\ $=2$ eV we obtain results in good agreement with the experimental transition temperature for the bulk. 
In addition, our results underline the importance of higher-order exchange-interactions, such as biquadratic exchange, in line with previous theoretical predictions for layered vdW materials \cite{kartsev20}.
Nevertheless, the spin-phonon coupling is most likely more pronounced in the single layer systems \cite{sadhukhan22} and our calculations also do not include the contribution of low-energy excitations, such as magnons.

We hope that our results can aid in the interpretation of future experiments on atomically thin VOCl layers, as well as the other members of the $M$OCl family, which also become distorted at low temperature, and which currently receive increasing attention \cite{miao18,zhang19,qing20,jang21,schaller22}.

\begin{acknowledgments}
We gratefully acknowledge financial support from Olle Engkvists stiftelse, grant 207-0582, and the Swedish e-Science Research Centre (SeRC).
All calculations were carried out using the facilities of the Swedish National Infrastructure of Computing (SNIC) at the National Supercomputer Centre (NSC), and the High Performance Computing Center North (HPC2N).
We thank Dr.\ L.\ Schoop for useful discussions.
The guidance provided by Dr.\ G.\ Bhilmayer, especially for the FLEUR calculations is gratefully acknowledged by the authors.
\end{acknowledgments}

\bibliographystyle{apsrev4-1}
\bibliography{bib}
\end{document}